\documentclass[11pt]{article}
\usepackage{geometry}          
\usepackage{graphicx}
\usepackage{amssymb}
\usepackage{epstopdf}
\usepackage{amsmath}
\usepackage{fixmath}
\usepackage{MnSymbol}
\usepackage{amsfonts}
\usepackage{bm}

\DeclareMathAlphabet{\mathpzc}{OT1}{pzc}{m}{it}

\title{The Superrotation of Venus: Where's the Torque?}
\author{Clifford Chafin\\\ \small{Department of Physics, North Carolina State University, Raleigh, NC 27695} \thanks{cechafin@ncsu.edu}}

\begin{document}
\maketitle

\begin{abstract}
The superrotation of the atmosphere of Venus requires a large torque on the upper atmosphere.  Mechanisms for providing a net balancing of this through waves or ionospheric motions to other parts of the atmosphere have been proposed but all have difficulties.  Here we demonstrate that the albedo gradient from the day to night side of the cloud layer allows a gradient of light pressure that is sufficient to provide an external torque to drive this flow.  
\end{abstract}
 
Venus is a peculiar planet in many respects.  It is roughly the size of the Earth yet has almost no magnetic field.  Despite this an being much hotter, it has held on to a much larger atmosphere than the earth. Its rotation rate about its axis is almost zero compared to other planets (equatorial surface velocity of 1.8 m/s) and it turns opposite to the direction of its orbit, retrograde, also unlike all the other planets.  Its rotational axis has very small declination and the vast majority of its surface is nearly featureless with over 50\% of its surface covered by variations less than 500m.  However, it is surrounded by dense atmosphere that rotates about the planet at $\sim100$m/s in the upper atmosphere and around 10m/s near the surface.  This rotation does not appear to be in the form a convection cells that give a net averaged motion of zero relative to the surface.  The upper atmosphere moves around the surface of the planet over about four (Earth) days and forms a pair of vortices at each pole.   This phenomenon is known as superrotation\footnote{``Superrotation'' in reference to the Earth refers to an unbroken continuous flow near the equator but this does not create any problems with conservation of angular momentum as there is back flow elsewhere.}.

There have been hydrodynamic and tidal models designed to try to generate and perpetuate this effect and the results are still controversial \cite{Gold, DelGenio}.  Recently these forces were investigated as a possible result of the transterminator flow of ions above the planet \cite{Durand}.  The forces between the ionosphere and the lower amplitude regions are supposed to be mediated by pressure and acoustic waves.  Two possible concerns arise here.  Firstly is the power provided in driving the transterminator flow.  It is suggested that this is from the solar wind.  The total force on Venus from the solar wind is $\sim10^{-5}$N so this has little capacity to produce much atmospheric motion.  Secondly, the use of waves to generate bulk forces can open up many subtle questions about the distinction between pseudomomentum and real momentum \cite{McIntyre}.  In the interior of an otherwise fixed medium, pseudomomentum is conserved and have units of momentum but not give corresponding forces on the walls of a container or the bulk motion of the medium as they dissipate.  As such, energy dissipation calculations may have little indication of the actual forces deposited on the atmosphere.  

\begin{figure}[!ht]
   \centering
   \includegraphics[width=2in,trim=0mm 0mm 0mm 0mm,clip]{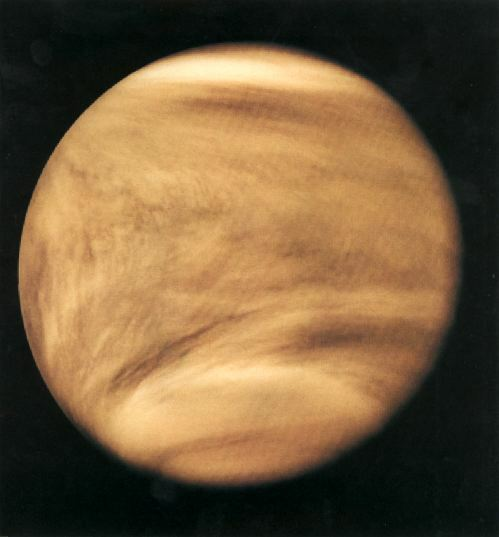} 
   \caption{Venus: clouds advance to the left.}
      \label{Venus}
\end{figure}

We argue here that the superrotation is due to an actual increase in the net angular momentum of the planet by photon momentum deposited at the level of of the opaque cloud layer at $\sim60$km above the surface.  The motion we observe is in the diurnal variation of the clouds not the underlying nearly nonrotating surface so a morning-evening albedo variation in the clouds is with respect to this notion of a Venusian day.  Such a variation is not unexpected.  At dawn the clouds have been cooling all night so we expect a more uniform layer.  By evening, the clouds have been subjected to solar heating for many hours and experienced the changing winds and convection driven by it.  Observations (see fig. \ref{Venus}) show that the clouds tend to be more uniform in the morning and more corrugated with dark spreading troughs in the evening.  we investigate the light pressure forces on these clouds let us start with an idealized example.  

\begin{figure}[!ht]
   \centering
   \includegraphics[width=2in,trim=0mm 0mm 0mm 0mm,clip]{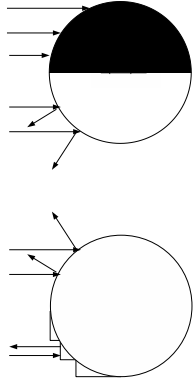} 
   \caption{Half black and half reflective (specular) vs.\ half reflective and half oriented normal reflection.  }
      \label{reflection}
\end{figure}

Consider a spherical body that is half dark and half light (like Iapetus) that is situated so that half of each received the sun's rays as in fig.~\ref{reflection}.  If the light side is polished to be (perfectly) reflective, then the rays are reflected specularly and the net momentum impulse is normal to the sphere's surface  and no net torque is imparted.  However, on the dark side (albedo $\approx 0$) the rays are absorbed imparting a torque.  The absorbed energy is then thermally radiated at a net normal to the surface imparting no torque of its own.  
The torque on the sphere is then 
\begin{align}
\tau=\int zdF=\int zP_{rad}dA=\int_{0}^{R} zP_{rad}(2\sqrt{R^{2}-z^{2}})dz=\frac{2}{3}P_{rad}R^{3}
\end{align}
Clouds will exhibit partially diffusive reflection which will give an intermediate result between these two results but the order of magnitude must be comparable to the dark sphere result.  

To estimate the torque on the atmosphere is a challenging problem.  The flow is clearly turbulent with the Reynolds number $Re=\frac{\rho v d}{\mu}\sim 10^{13}$.  If we were to model this with the Darcy-Weisbach equation for turbulent flow in a pipe 
\begin{align}
\delta P=f_{D}\frac{L}{D} \frac{\rho v^{2}}{2}
\end{align}
where we use the pipe diameter, $D$, as the atmospheric depth $h=65$km and then break up the area of the planet $4\pi R^{2}=4.5\times10^{14}$m$^{2}$ into $\pi R/d\approx300$ tubes.  The Darcy friction factor, $f_{D}\approx0.04$ can be found from a Moody table as a function of the elevation of the surface roughness and the tube diameter.   We then obtain a net force on the atmosphere to be $\sim10^{19}$N.  The solar illumination pressure at Venus is $P_{rad}\approx 9\times10^{-6}$Pa so the net force on the planet is $\sim10^{9}$N so clearly this is much too small.  

Modeling turbulence is a difficult and very geometric dependent problem and this order of magnitude calculation turns out to be very wrong.  Assuming the flow is a stationary state, all the traction on the clouds is delivered to the planet where the wind velocity is measured by probes to be $\lesssim1$m/s.  Assuming the height of these measurements is about a meter, we have the net force of $F< \mu \frac{dv}{dy}A\approx 10^{10}$N where the viscosity of CO$_{2}$ is $\eta=2.5\times10^{-5}$Pa$\cdot$s.  This rough calculation gives a force of similar size to that corresponding to the maximum torque that the light pressure could apply. Our information about the net forces present on the surface are limited by the small number of probes that have landed there and their relatively short lifespan under such extreme conditions.  

We can get an estimate of the albedo gradient from photographs of the bright side of Venus.  The atmosphere is very changeable and we would ideally like to get a sense of the absorption, radiation, and reflection over all angles at all locations of the surface.  Since we are only seeking an order of magnitude estimate here let us analyze a somewhat typical image in fig. \ref{Venus}.  

\begin{figure}[!ht]
   \centering
   \includegraphics[width=2in,trim=0mm 0mm 0mm 0mm,clip]{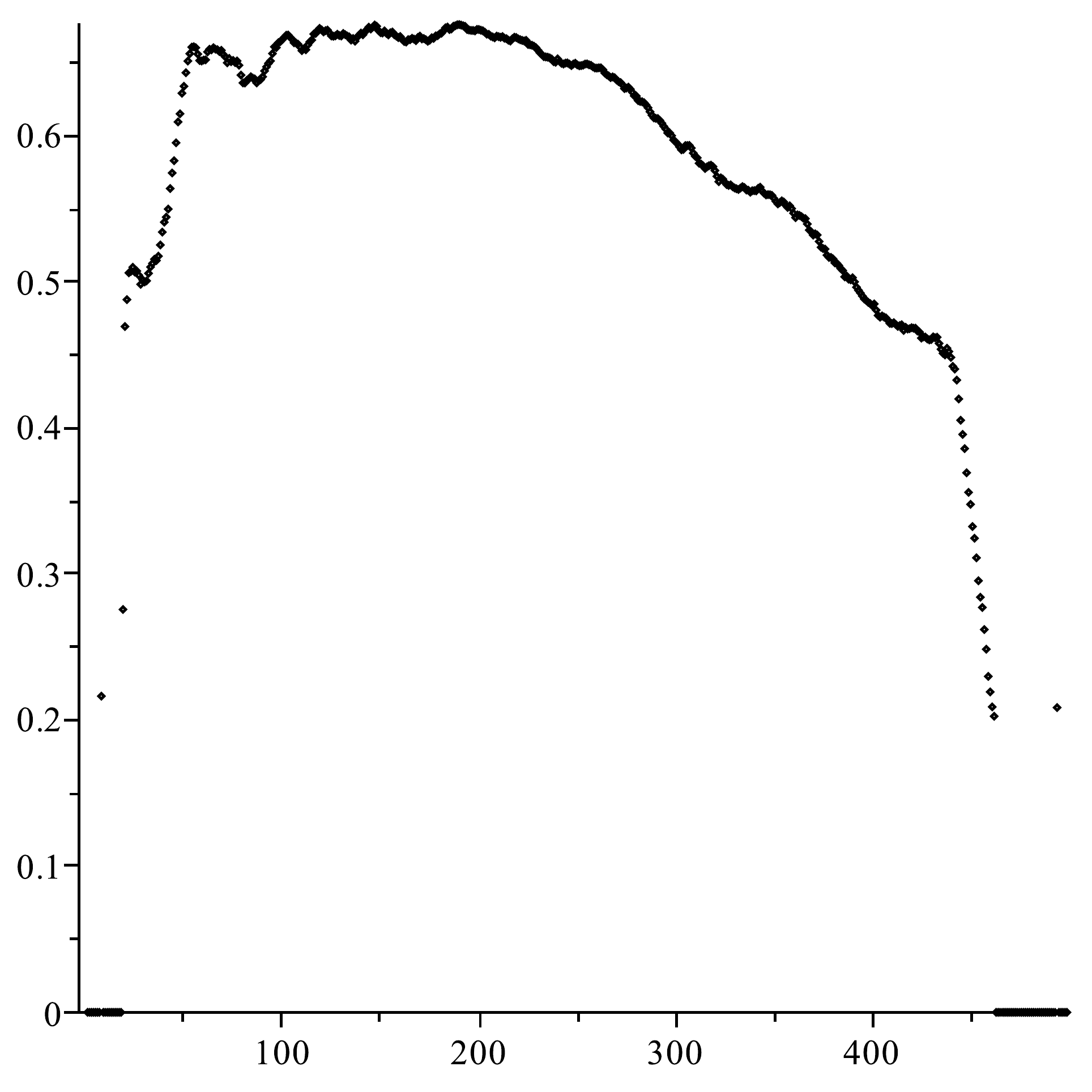} 
   \caption{The vertically averaged brightness across the image as a function of pixel width.}
      \label{brightness}
\end{figure}

\begin{figure}[!ht]
   \centering
   \includegraphics[width=2in,trim=0mm 0mm 0mm 0mm,clip]{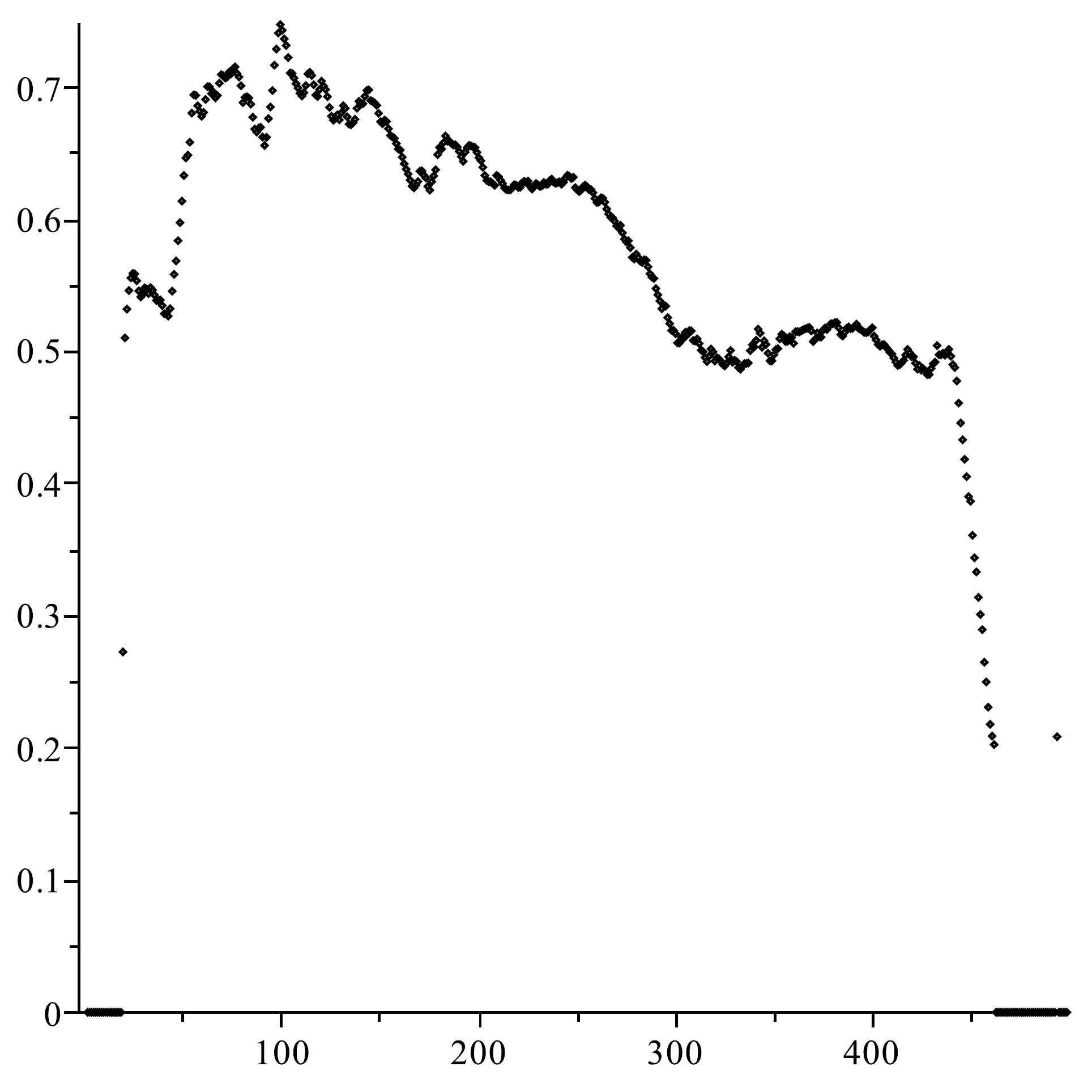} 
   \caption{The vertically averaged brightness across the middle 40 pixels about the equator.}
      \label{middle}
\end{figure}

The dark cloud fissures on the evening side of Venus will act to absorb radiative energy and momentum which then is thermally radiated normally to the planet.  Furthermore, the angled surfaces of the clouds breaking up on this side can direct diffusive reflection more directly backwards.  These contribute to  a small but visibly discernible albedo change across its cloud face.  We see this from the integrated brightness profiles of Venus\footnote{\text{http://nssdc.gsfc.nasa.gov/image/planetary/venus/pvo\_uv\_790205.jpg} Pioneer orbiter 1979}  fig.~\ref{Venus} in Figs.~\ref{brightness} and \ref{middle}.  
From the fraction of this we can reduce the above maximal torque to the actual value.  In our naive straight-on integrated brightness profile we see a 36\% increase in brightness on the evening side over the morning side of the planet.  

This suggests that the pressure of light is playing a strong role in the clouds behavior and may be providing the vast majority of the torque on the clouds or stabilizing it against perturbations.  One should not try to read to much into this value.  If we return to our example of a half white, half black body the the torque can be completely cancelled if we have the ``white'' side change from reflecting specularly to diffusively.  If the reflections are from  surfaces that not normal to the planet are to the incident rays this can completely reverse the torque.

The complexity of the system is so rich, involving turbulence and details in daily specular properties over many frequencies that involve droplet evolution that it is unclear that any numerical simulations would give much of an authoritative picture on the system.  The best way forwards is likely to do a detailed broad spectrum study of the electromagnetic emissions over all angles from Venus then to build a larger database of surface wind measurements to give a comparison of solar torque versus large scale hydrodynamic response.  

Given the similar size and distance to the sun of Earth and Venus, one cannot help to wonder if similar forces exist on our own planet.  There is certainly a diurnal cycle of cloud variation based on solar heating, wind variation and evaporation.  The low viscosity of air and the long distances between the clouds and the surface allow such relatively low forces to transform the modest torques on the upper atmosphere to significant accelerations and directional biases of motion.  In this age of great interest and investment in climate modeling where even cosmic rays seem to affect the weather, it would be vital and amusing to find that such a puny cosmic force as the pressure of light was shaping our planet as well.

\end{document}